\documentclass[twocolumn]{aastex63}

\usepackage{etoolbox}
\usepackage{graphicx}	
\usepackage{multirow}
\usepackage{cleveref}
\usepackage{CJK}
\usepackage{comment}
\newcommand{\cloudy}{\textsc{cloudy}}
\newcommand{\kms}{km s$^{-1}$}
\newcommand{\cmN}{cm$^{-2}$}

\newcommand{\lam}{$\lambda$}

\newcommand{\lya}{\mbox{Ly$\alpha$}}

\newcommand{\lye}{\mbox{Ly$\epsilon$}}
\newcommand{\hi}{\mbox{H\,{\sc i}}}

\newcommand{\civ}{\mbox{C\,{\sc iv}}}

\newcommand{\siiv}{\mbox{Si\,{\sc iv}}}

\newcommand{\nv}{\mbox{N\,{\sc v}}}

\newcommand{\ovi}{\mbox{O\,{\sc vi}}}

\newcommand{\target}{J151352+085555}

\received{}
\revised{}
\accepted{}
\submitjournal{ApJ}
\shorttitle{Line-locking from QSO \target }
\shortauthors{Chen et al.}

\begin{document}

\title{Tracking Outflow using Line-Locking (TOLL). II. Large Line-Locking Web identified in Quasar~\target}

\correspondingauthor{Bo Ma}
\email{mabo8@mail.sysu.edu.cn}

\author{Chen Chen}
\affiliation{Zhuhai College of Science and Technology, Zhuhai 519000, China}

\author{Zhicheng He}
\affiliation{CAS Key Laboratory for Research in Galaxies and Cosmology, Department of Astronomy, University of Science and Technology of China, Hefei 230026, China}

\author{Weimin Yi}
\affiliation{Yunnan Observatories, Kunming 650216, China}

\author{Tuo Ji}
\affiliation{Center for Space Physics and Astronomy, Key Laboratory for Polar Science of the State Oceanic Administration, Polar Research Institute of China, Shanghai 200136, People's Republic of China}

\author{Marie Wingyee Lau}
\affiliation{Department of Physics and Astronomy, University of California Riverside, Riverside, CA 92507, USA}

\author{Bo Ma}
\affiliation{School of Physics $\&$ Astronomy, Sun Yat-Sen University, Zhuhai 519000, China}
\affiliation{CSST Science Center in the Great Bay Area, Sun Yat-sen University, Zhuhai 519000, China}

\begin{abstract}

Quasar outflows often consist of two clouds with velocity separations matching the doublet spacings of common UV resonance transitions, a phenomenon known as line-locking,  which is commonly observed in quasar spectra. 
Multiple clouds can be locked together through multi-ion doublets, forming `line-locking web'. 
In the second paper of the TOLL project, we present discovery of one of the largest `line-locking web' known to date from the VLT/UVES spectra of QSO~\target. 
We identify 12 associated narrow absorption line systems through the \civ, \nv, \siiv, \ovi, and multiple Lyman lines (\lya\ to \lye), and find 10 out of the 12 absorbers are line-locked together by comparing velocity separations between different absorption systems.
By conducting photoionization modeling with CLOUDY, we measure the total hydrogen column densities, metallicities, and ionization parameters of these absorbers, which suggests the absorbers likely have sub-solar metallicities.
A preliminary statistical analysis demonstrates the shadowed clouds tend to have similar ionization states comparing to the shadowing ones. 
Identification of one of the largest line-locking webs implies that radiative acceleration plays an important role in sorting out cloud velocities in quasar outflows, and highlights the need for more sophisticated theoretical models to explain its formation and evolution.
\end{abstract}

\section{Introduction}
Quasar outflows play a crucial role in galaxy evolution by providing kinetic energy feedback from central supermassive black holes to their host galaxies \citep{Silk98, Kauffmann00, Berti08, Costa14, Barai16, Angl17a, Angl17, Barai19, Bustamante19, He22, Chen22, Bluck23, Ayubinia23, Naddaf23, Hall24, He24, Bollati24}. 
Several possible outflow/wind driving mechanisms, including thermal process, radiation pressure, and magnetic fields, have been investigated extensively \citep{Mushotzky72, Scargle73, Begelman83, Krolik86, Camenzind86, Pelletier92, Netzer93, Arav94b, Arav95, Kool95, Murray95, Proga00, Proga04, Chelouche05, Netzer06, Proga07b, Baskin12, Baskin14, Netzer24}.
One common and distinctive feature observed in these outflows, which is often considered as evidence supporting the radiation pressure mechanism, is line-locking \citep{Milne26, Scargle73, Burbidge75, Goldreich76, Foltz87, Braun89}.
It is a phenomenon where the velocity difference between two different absorption components matches the separation of known atomic doublet transitions, like the \civ\ doublets \citep{Foltz87, Srianand02, Ganguly03, Gallagher04, Hamann11, Bowler14, Chen19, Lin20a, Lin20b, Lu20, Chen24}. 
In the geometrical configuration, the slower (shadowing) component lies between the emission source and the other faster (shadowed) outflow component.
This configuration will be maintained as long as the acceleration of the shadowed component via radiation pressure through the red doublet stays equal to or greater than what the acceleration difference between the two outflow components would be outside of the shadow. 
Thus, studying line-locking can provide valuable insights into the role of radiation pressure played in driving and shaping quasar outflows.

The formation and evolution of line-locking between two absorbing clouds through two resonance lines in active galaxies has been explored theoretically. 
For instance, \citet{Braun89} proposed a popular non-steady state mechanism to explain the formation of line-locking. \citet{Korista93} provided a detailed analysis of the line lock theory in BALQSOs, suggesting that radiation pressure and line-locking mayor play crucial roles in the formation of the observed double troughs in the \civ\ 1549 region, and has significant implications for understanding the dynamics of BAL outflows. 
\citet{Arav94c} provided a dynamical model that links the observed double troughs in the \civ\ BAL of BALQSOs to the modulation of the radiative force due to line locking between \lya\ and \nv~$\lambda$~1240 resonance lines \citep[see also][]{Arav96}. 
\citet{Proga00} extended previous models of line-driven winds, originally developed for hot stars, to the axially symmetric geometry of accretion disks in AGN. 
\citet{Vilkoviskij01} showed that theories based on
small cloud velocity interactions via line locking can explain the spectral observation of the BALQSO Q1303+1308.
\citet{Chelouche03} modeled the impact of shielding on properties of outflows, and found moderate shielding is especially efficient in accelerating flows to high velocities because of the suppression of gas ionization level.
\citet{Baskin12} utilized a `metal ion runaway' model to explain the apparent lack of \lya-\nv\ line-locking signature in \civ\ absorption troughs of some AGNs. 
\citet{Lewis23} studied a simplified two-clouds-one-doublet model and found that fine-tuning of physical parameters such as column density and ionization is required for line-locking to occur.

However, in practice, multiple clouds have been observed to be locked together through multi-ion doublets rather than just two clouds, and whether fine-tuning is needed in such cases remains questionable \citep{Chen24}.
For example, \citet{Hamann11} detected 3 narrow absorption line systems (NALs) toward a quasar, line-locked together by the \civ\ doublet \citep[see also][]{Lin20a, Lin20b, Lu20, Chen21, Yi24, Chen24}. 
For simplicity, we hereafter use the new concept of `line-locking web' to describe the structure of the interconnected line-locking clouds in a quasar outflow.
To study the formation and evolution of the line-locking web and develop more realistic theoretical models, we have initialed the Tracking Outflow using Line-Locking (TOLL) project \citep[][hereafter Paper I]{Chen24}. 
The primary goal of the TOLL project is to first establish a well-characterized sample of line-locked systems using high resolution quasar spectra, and then conduct statistical and theoretical work based on the sample in the future. 
In the second paper of the TOLL project, we present identification of one of the largest line-locking web using high-resolution spectra of quasar~\target\ and conduct a statistical test of the `fine-tuning' scenario proposed by \citet{Lewis23} using data from this and our previous works \citep{chen18, Chen19, Chen24}. 

The outflow of quasar~\target\ was investigated by \citet{Srianand02}, where they have identified line-locking signatures corresponding to the \ovi, \nv, and \civ\ doublet splittings. 
In this paper, we re-analyze the UVES spectra of quasar~\target\ and report new findings in terms of its line-locking signatures. 
The structure of the paper is organized as follows. 
Section~2 introduces the UVES spectrum of \target\ and the methodology for fitting line profiles of associated absorption lines \citep[AALs;][]{Weymann79, Foltz86, Anderson87, Weymann91, Hamann97d}, used to measure the Doppler parameters, column densities, velocity shifts, and covering fractions of each absorber. 
In Section~3, we describe results from line fitting, line-lock searching, and photoionization modeling. 
Section~4 compares our findings with results of \citet{Srianand02}, and discusses the implications of our results. 
Finally, Section~5 provides a summary of this work. 

Throughout this paper, we adopt a cosmological model with $H_0 = 71$ \kms\ Mpc$^{-1}$, $\Omega_M = 0.27$, and $\Omega_\Lambda = 0.73$.

\section{QUASAR SPECTRA and DATA ANALYSIS}

\subsection{Quasar Spectra}
The TOLL project’s parent sample of quasars is based on the catalog compiled by \citet{chen20}, which includes observations from VLT-UVES \citep{Murphy19} and Keck-HIRES \citep{OMeara15}. These spectra have resolutions ranging from \(22,000 \lesssim R \lesssim 71,000\) for VLT-UVES. In the \citet{chen20} study, each candidate mini-broad absorption line (mini-BAL) system was carefully examined to ensure it was not contaminated by unrelated intervening absorption features, such as damped \lya\ or Lyman limit systems originating from foreground galaxies or their extended halos \citep[e.g.,][]{Prochaska15, Berg15, Prochaska08}. 
Our main goal is to study these line-locked features in order to gain deeper insights into the dynamics of quasar outflows. 
Thus, we conducted a thorough visual inspection of both the normalized and unnormalized spectra of all quasars in the catalog, and selected \target\ for its promising potential to reveal line-locked systems.

Quasar \target\ (also named Q1511+091, SDSS~J151352.52+085555.7) is optically bright ($m_{\rm B} = 18.0$) and has a emission redshift of $z=2.901$ measured from SDSS \citep{Abolfathi18, Paris18}. 
It is classified as a \civ\ broad absorption line (BAL) quasar, and has a bolometric luminosity of $10^{14}L_\odot$ and virial black hole mass of $2.5\times10^{10}M_\odot$ \citep{Shen11, Weedman12}.
\citet{Sargent88} studied its \civ\ absorption using the Palomar 5m Hale Telescope, and found a strong complex absorption line system at a absorption redshift of $z_{obs} \sim 2.85$. 
It was found to be radio intermediate from the VLA FIRST survey, having a radio loudness of 1.34, and was given the name of FIRST~J151352.5+085554 \citep{Helfand15, Shaban22}.
A point source, CXOG J151352.5+085555, was detected in the X-ray band by Chandra near the optical position of quasar~\target, with an variability amplitude about 50\% and an average X-ray luminosity of $\rm{4\times10^{45}~erg~s^{-1}}$ \citep[0.5-7~kev,][]{Wang16, Shaban23, Quintin24}. 

\begin{figure*}
\centering
\includegraphics[width=1.0\textwidth]{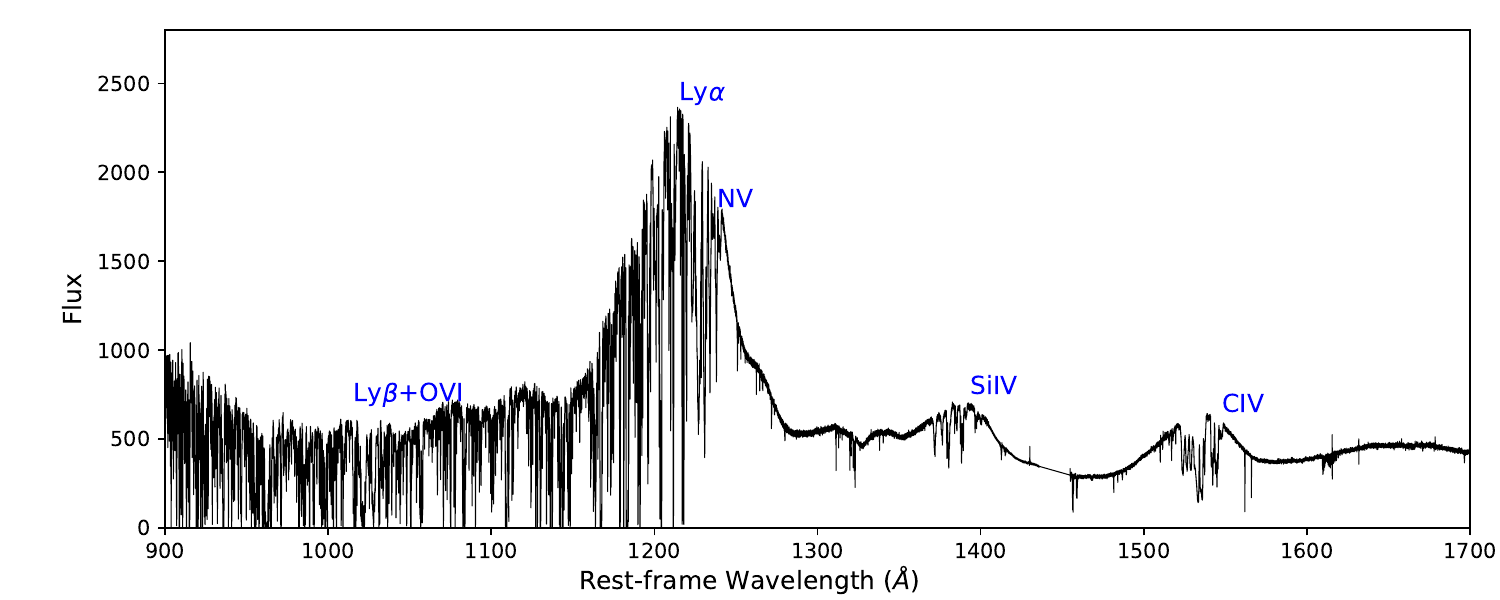}
\caption{UVES spectrum of \target\ at the rest-frame wavelengths (defined by the emission redshift $z=2.901$) showing the absorption lines relative to the broad emission lines. The main emission lines are labeled across the top.\label{fig:spectrum}}
\end{figure*}

UVES is a dual-arm, grating cross-dispersed high-resolution optical echelle spectrograph mounted on the Nasmyth B focus of Unit Telescope 2 of the VLT \citep{Dekker00}. 
The UVES observations of quasar~\target\ (PI: Srianand; ProgramID: 65.P-0038, 69.B-0108, 71.B-0136) were conducted in 2000, 2002 and 2003 under a medium seeing of 0.56~arcsec, with a total on-target exposure time of 52713~seconds \citep[see also][]{Murphy19}.  
Seventeen exposures were taken in the UVES standard configuration using the 0.8'' and 1'' slit, providing a spectral resolution of R$\sim$50,000 and wavelength coverage of 3284$-$10429~\AA.
\citet{Murphy19} reduced the data using a modified version of the UVES reduction pipeline based on the optimally extracted method \citep{Horne86, Piskunov02, Zechmeister14, Ma19}. 
The wavelength calibration is done using a single ThAr exposure \citep{Murphy07}. 
We show the target spectrum at the rest-frame wavelengths in \Cref{fig:spectrum}.
We did not detect significant time-variable absorption lines from the spectra taken about three years apart.
Different observations then were combined to form a single spectrum with a dispersion of 2.5~\kms per pixel. 
The final continuum-normalized spectrum has a count-to-noise ratio of 19, 40, 77, 78, and 52 per pixel (2.5~\kms) near 3500, 4500, 5500, 6500, and 7500~\AA\ in the observed frame. 

\subsection{Line Profile Fitting}
We use a standard fitting procedure to extract all relevant absorption line systems from the quasar spectra \citep{chen18, Chen19, chen20}.
The fitting process begins with a thorough examination of the continuum placement provided in the UVES archive spectra for each line of interest. 
In cases where the continuum appears to be poorly constrained or exhibits sharp variations compared to the unnormalized spectrum, we refit the continuum using a locally-constrained simple power law.
Then we start to identify individual velocity components in the associated absorption line (AAL) complexes, beginning with the \civ\ \lam 1548, 1551 doublet. 
This doublet is typically strong in our target spectra, occupies wavelength range outside of the Lyman forest, and has a high signal-to-noise ratio. 
We then search for a wide range of other plausible lines at the corresponding redshifts to the \civ\ doublets, such as \siiv\ \lam 1394, 1403, \nv\ \lam 1239, 1243, \ovi\ \lam 1032, 1038, and the Lyman series. Upon identifying new components at the redshifts associated with other absorption lines, we proceeded to verify the presence of corresponding \civ\ lines in return. This process of mutual validation continued until we had thoroughly cataloged all the lines involved.
Our spectra include the AAL complexes of \civ, \siiv, \ovi, and \nv\ doublets. 
In some cases, other lines help confirm the positions of \civ\ components. For instance, component 5 in \Cref{fig:civ} are identified through \siiv.
\Cref{tab:J1513} lists all identified lines in the quasar spectra.

Our line-fitting procedure follows the approach detailed in \citet{chen18, Chen19}. We fit each candidate AAL for \civ, \siiv, \ovi, \nv\ and the Lyman series using a Gaussian optical depth profile:
\begin{equation}
\label{eq:1}
\tau_{v} = \tau_{0} \exp \left( -\frac{(v - v_0)^2}{b^2} \right),
\end{equation}
where $\tau_v$ is the optical depth at velocity $v$, $\tau_0$ is the central optical depth, $v_0$ is the line center velocity, and $b$ is the Doppler parameter. The two lines in each doublet are fit simultaneously, sharing the same velocity shift and $b$ value. This procedure can also apply to the multiple lines in Lyman series, as they also share the same velocity shift and $b$ value.

We account for partial coverage of the background light source in the fitting, assuming a spatially uniform brightness of the source and a homogeneous absorbing medium. The observed intensity at velocity $v$ is then expressed as:
\begin{equation}
\label{eq:2}
\frac{I_{v}}{I_{0}} = 1 - C_0 + C_0 \exp(-\tau_{v}),
\end{equation}
where $I_{0}$ is the continuum intensity, $I_{v}$ is the measured intensity at velocity $v$, and $C_0$ is the covering fraction along our line of sight, with $0 < C_0 \leq 1$ \citep{Ganguly99, Hamann97b, Barlow97b}. We assume a constant covering fraction across the AAL profiles, attributed to the narrowness of the lines and the minimal variation in their lower extremities. But the value of $C_0$ can differ between lines.

The column densities for \civ, \siiv\, \ovi, and \nv\ are derived from the fitted optical depths using the relation:
\begin{equation}
\label{eq:3}
\tau_0 = \frac{\sqrt{\pi} e^2}{m_e c} \frac{N_i f_{\mathrm{oscillator}} \lambda_0}{b},
\end{equation}
where $N_i$ is the ion column density, $f_{\mathrm{oscillator}}$ is the oscillator strength, $\lambda_0$ is the rest wavelength, and $b$ is the Doppler parameter.

While most fits are straightforward, certain lines present challenges due to blending with other absorption systems or unrelated lines. When multiple systems are blended, we fit the doublets simultaneously and resolve individual AALs only if they can be clearly distinguished visually. For regions containing unrelated absorption at different redshifts, we mask these areas before conduct the fitting procedure. Most \ovi\ lines are heavily contaminated within the \lya\ to \lye\ forests. We estimate their lower limits for conservation. For the components that show multiple Lyman lines, we try to fit as many Lyman lines as possible simultaneously to get the best constraints on the column densities and covering fractions.

\section{Results}
\subsection{Line Fitting Parameters}
\begin{figure*}
\centering
\includegraphics[width=0.9\textwidth]{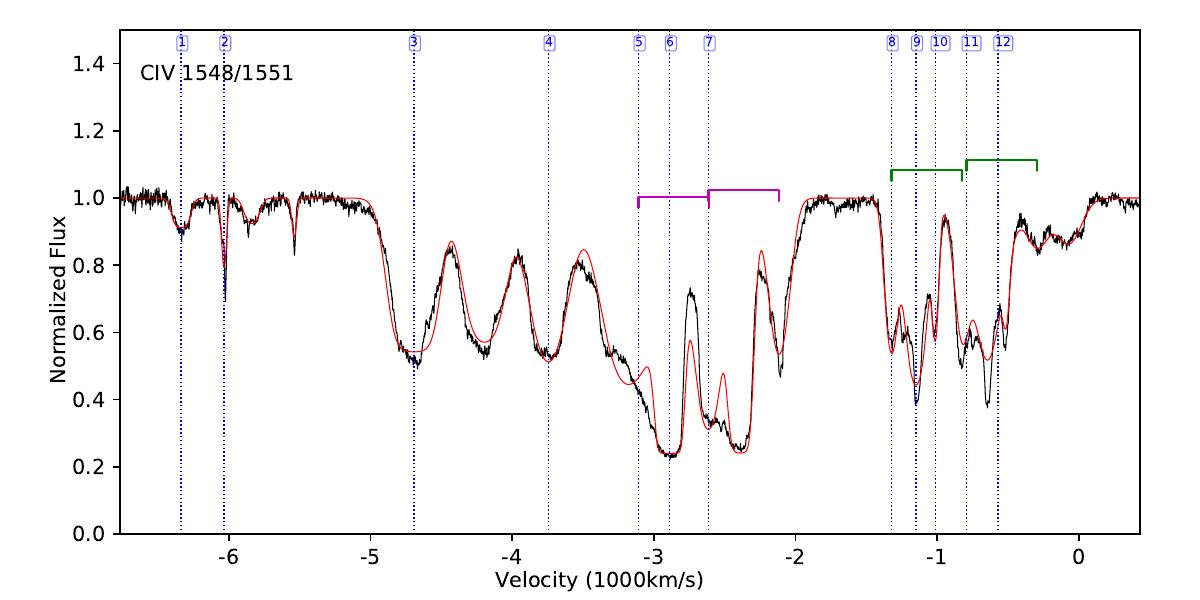}
\caption{Normalized \civ\ line profiles in the VLT-UVES spectra plotted on a velocity scale relative to the quasar redshift $z=2.901$. The spectra are shown in black, and the final fitting lines are shown in red. The blue dash lines are identified components from 1 to 12, and the brackets show the line-locked doublets. The velocities pertain to the short-wavelength lines in the doublets.\label{fig:civ}}
\end{figure*}

\begin{figure*}
\centering
\includegraphics[width=0.9\textwidth]{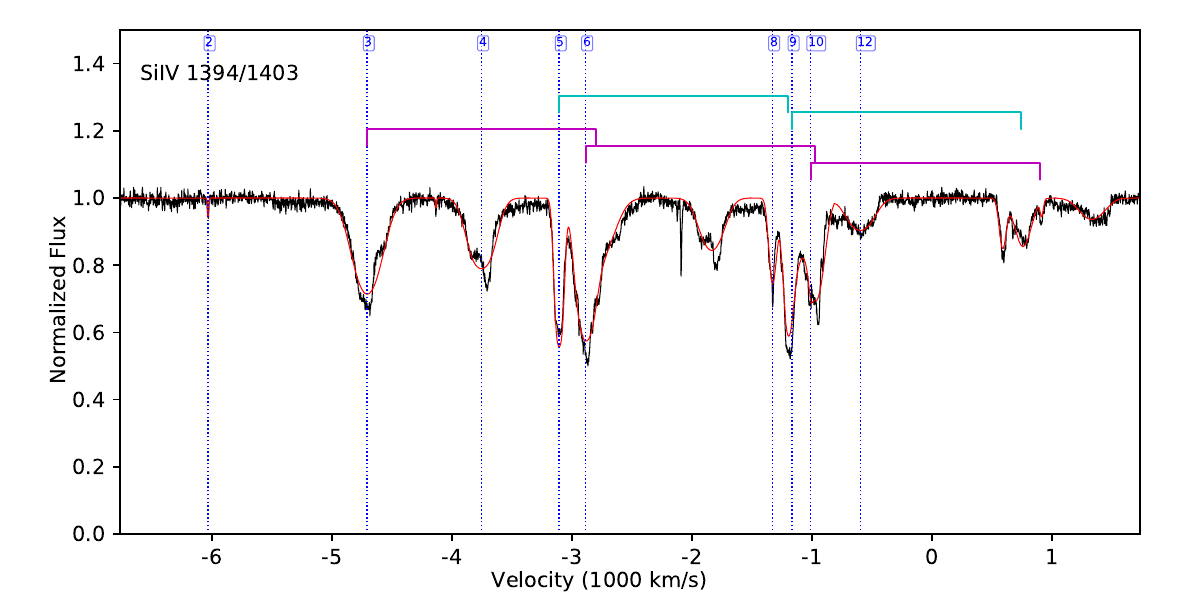}
\caption{Normalized \siiv\ line profiles in the VLT-UVES spectra plotted on a velocity scale relative to the quasar redshift $z=2.901$. The spectra are shown in black, and the final fitting lines are shown in red. The blue dash lines are identified components, and the brackets show the line-locked doublets. The velocities pertain to the short-wavelength lines in the doublets.\label{fig:siiv}}
\end{figure*}

\begin{figure*}
\centering
\includegraphics[width=0.9\textwidth]{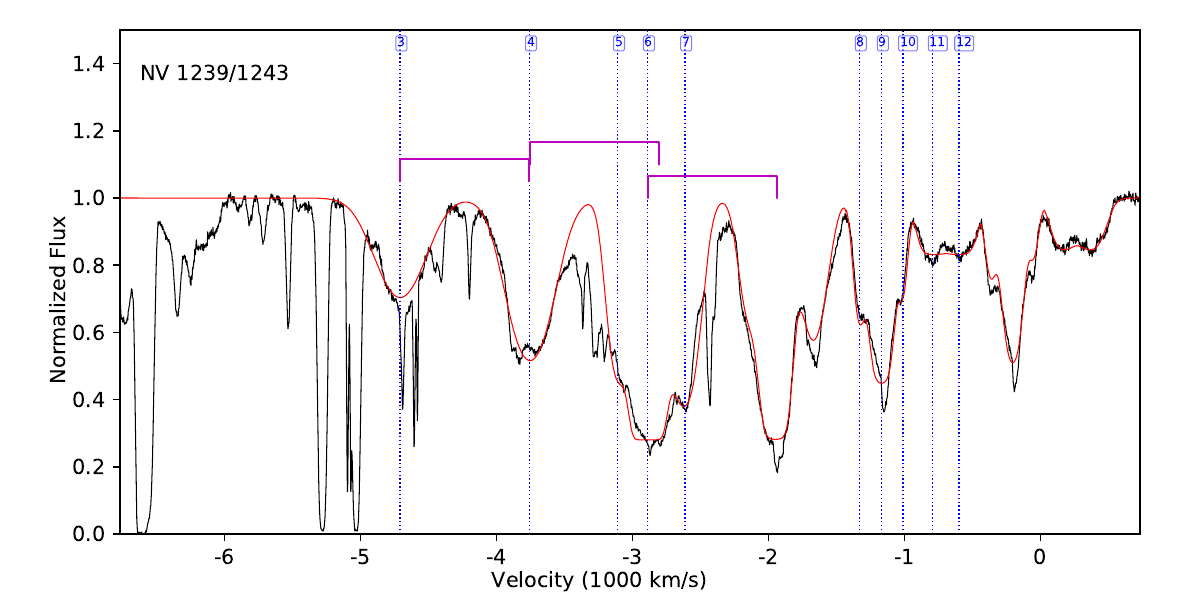}
\caption{Normalized \nv\ line profiles in the VLT-UVES spectra plotted on a velocity scale relative to the quasar redshift $z=2.901$. The spectra are shown in black, and the final fitting lines are shown in red. The blue dash lines are identified components, and the brackets show the line-locked doublets. The velocities pertain to the short-wavelength lines in the doublets.\label{fig:nv}}
\end{figure*}

\begin{figure*}
\centering
\includegraphics[width=0.9\textwidth]{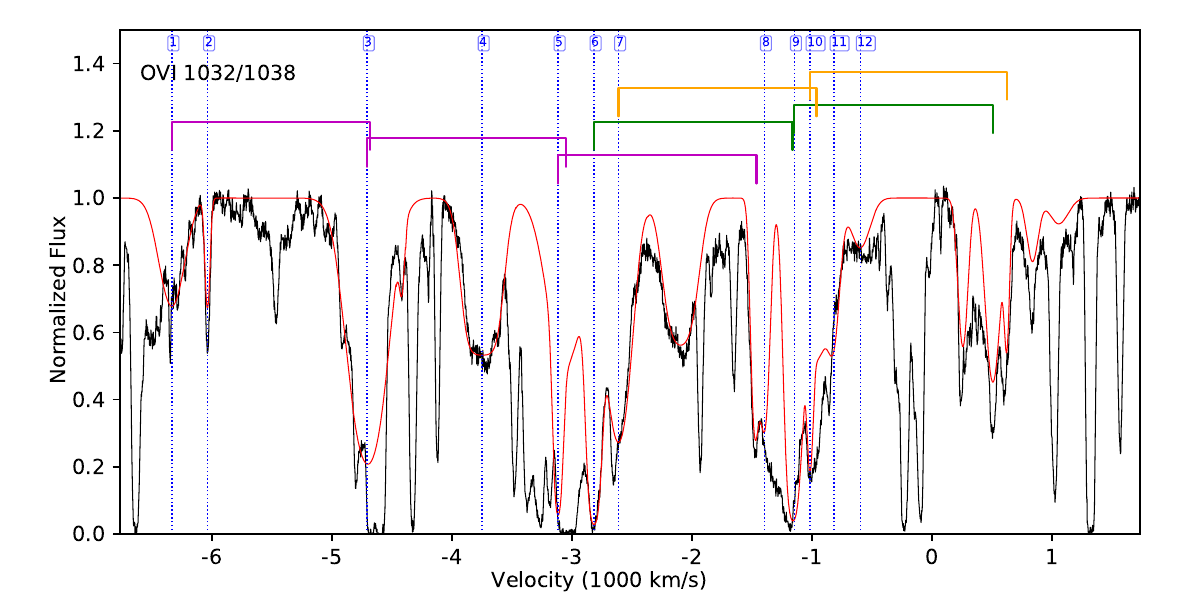}
\caption{Normalized \ovi\ line profiles in the VLT-UVES spectra plotted on a velocity scale relative to the quasar redshift $z=2.901$. The spectra are shown in black, and the final fitting lines are shown in red. The blue dash lines are identified components, and the brackets show the line-locked doublets. The velocities pertain to the short-wavelength lines in the doublets.\label{fig:ovi}}
\end{figure*}

\begin{figure*}
\centering
\includegraphics[width=0.9\textwidth]{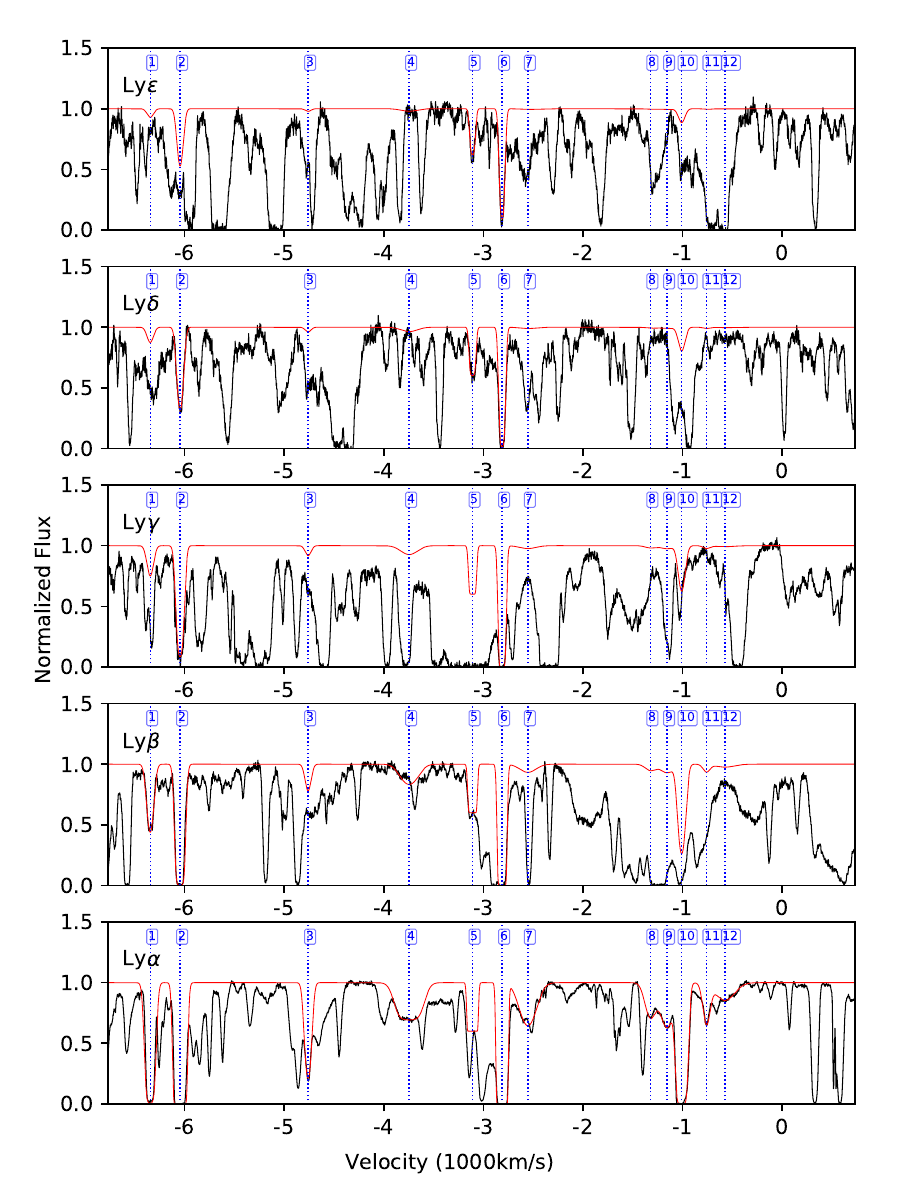}
\caption{Normalized Lyman line profiles in the VLT-UVES spectra plotted on a velocity scale relative to the quasar redshift $z=2.901$. The spectra are shown in black, and the final fitting lines are shown in red. The blue dash lines are identified components from 1 to 12, and the brackets show the line-locked doublets. The velocities pertain to the short-wavelength lines in the doublets.\label{fig:Lyman}}
\end{figure*}

The fitted line parameters and corresponding uncertainties of the \civ, \siiv, \ovi, and \nv\ doublets from the VLT-UVES spectrum of quasar \target\ are summarized in \Cref{tab:J1513}. 
The table presents the measured quantities for each line, including the velocity shift $v$, line identification, rest wavelength, observed wavelength, Doppler parameter $b$, logarithm of the column density, and covering fraction $C_0$. 
These measurements are provided separately for the \civ, \siiv\, \ovi, and \nv\ doublets. 
The last column contains notes regarding any line blends. 
The data are organized by redshift, corresponding to the component numbers listed in the first column, as shown in \Cref{fig:civ,fig:siiv,fig:nv,fig:ovi,fig:Lyman}.
The uncertainties reported for most parameters represent the $1\sigma$ errors obtained from the fitting procedure, primarily reflecting pixel-to-pixel noise in the spectra. 
For cases of significant blending, we employ the methods outlined in Section 3.2 of \citet{chen18} and Section 3.1 of \citet{Chen19} to directly estimate parameter values or place limits on them.

\Cref{fig:civ} displays our line profile fits for all \civ\ doublets included in the final catalog. The velocity shift $v$, covering fraction $C_0$, and column density $\log N$ are derived from \Cref{eq:1,eq:2,eq:3} through fitting the \civ\ doublet line profiles. 
\Cref{fig:siiv}, \Cref{fig:nv},  \Cref{fig:ovi} and \Cref{fig:Lyman} illustrate the fitted profiles for the \siiv, \nv, \ovi, and Lyman AALs, respectively, as included in the final catalog.

\begin{deluxetable*}{ccccccccccc}
\tablecaption{Individual absorption lines of \target. Columns show component number, absorption redshift ($z_{abs}$) and the corresponding velocity shift ($\textrm{v}$) relative to the emission-line redshift 2.901 \citep{Murphy19}, line identification and rest wavelength, observation wavelength, Doppler b parameter, logarithm of column density, covering fraction, and notes ('bl'=blended with neighboring systems or unrelated lines, if the blended system is also measured, we record bl-X to indicate the component number of the other system.\label{tab:J1513}}
\tabletypesize{\scriptsize}
\tablehead{
\colhead{$\#$} & \colhead{$z_{abs}$} & \colhead{$\textrm{v}$ (\kms)} & \colhead{ID1} & \colhead{$\lambda_{obs1}$ (\AA)} & \colhead{ID2} & \colhead{$\lambda_{obs2}$ (\AA)} & \colhead{$b$ (\kms)} & \colhead{$\log N$ (\cmN)}  & \colhead{$C_0$} & \colhead{Notes} 
}  
\startdata
1 & 2.8186 & -6337 & \multicolumn{4}{c}{Lyman Series} & $44.0\pm1.2$ & $14.48\pm0.01$ & 1.0 & \\
 &  & & \ovi\ 1032 & 3940.51 & \ovi\ 1038 & 3962.24 & $129.6\pm10.3$ & $\gtrsim14.40$ & $0.33\sim1.0$ & bl,bl-3 \\
 &  &  & \civ\ 1548 & 5911.94 & \civ\ 1551 & 5921.77 & $54.4\pm2.0$ & $14.66\pm0.06$ & $0.09\pm0.01$ & \\
2 & 2.8226 & -6029 & \multicolumn{4}{c}{Lyman Series} & $39.3\pm0.7$ & $15.36\pm0.01$ & 1.0 & \\
 & & & \ovi\ 1032 & 3944.64 & \ovi\ 1038 & 3966.39 & $26.4\pm1.0$ & $14.25\pm0.08$ & $0.58\pm0.07$ & \\
 & & & \siiv\ 1394 & 5327.94 & \siiv\ 1403 & 5362.23 & $7.8\pm7.5$ & $11.64\pm0.36$ & 1.0 & weak line\\ 
 &  &  & \civ\ 1548 & 5918.13 & \civ\ 1551 & 5927.97 & $22.8\pm1.3$ & $13.41\pm0.31$ & $0.53\pm0.32$ & \\
3 & 2.8398 & -4706 & \multicolumn{4}{c}{Lyman Series} & $39.7\pm0.9$ & $\gtrsim13.91$ & $0.83\sim1.0$ & \\
 & & & \ovi\ 1032 & 3962.39 & \ovi\ 1038 & 3984.24 & $214.2\pm18.9$ & $\gtrsim15.15$ & $\leq1.0$ & bl,bl-1,5 \\
 &  &  & \nv\ 1239 & 4756.82 & \nv\ 1243 & 4772.12 & $240.0\pm18.9$ & $14.64\pm0.03$ & $0.72\pm0.05$ & bl,bl-4\\ 
 & & & \siiv\ 1394 & 5351.91 & \siiv\ 1403 & 5386.36 & $153.5\pm37.2$ & $14.12\pm0.11$ & $0.45\pm0.08$ & bl-6\\ 
 &  &  & \civ\ 1548 & 5944.76 & \civ\ 1551 & 5954.65 & $140.9\pm1.6$ & $15.24\pm0.02$ & $0.46\pm0.01$ & \\
4 & 2.8522 & -3753 & \multicolumn{4}{c}{Lyman Series} & $111.4\pm2.5$ & $14.85\pm0.02$ & $0.31\pm0.01$ & bl \\
 & & & \ovi\ 1032 & 3975.19 & \ovi\ 1038 & 3997.11 & $130.7\pm2.7$ & $15.54\pm0.03$ & $0.47\pm0.01$ &  \\
 &  &  & \nv\ 1239 & 4772.19 & \nv\ 1243 & 4787.53 & $197.9\pm6.1$ & $14.76\pm0.10$ & $0.46\pm0.19$ & bl-3,5,6,7\\ 
 & & & \siiv\ 1394 & 5369.20 & \siiv\ 1403 & 5403.75 & $120.8\pm3.0$ & $14.37\pm0.04$ & $0.24\pm0.02$ & \\ 
 &  &  & \civ\ 1548 & 5963.96 & \civ\ 1551 & 5973.88 & $164.5\pm1.7$ & $14.58\pm0.05$ & $0.57\pm0.08$ & bl-5\\
5 & 2.8606 & -3107 & \multicolumn{4}{c}{Lyman Series} & $25.8\pm3.0$ & $\lesssim15.87$ & $0.40\sim1.0$ & bl \\
 & & & \ovi\ 1032 & 3983.85 & \ovi\ 1038 & 4005.82 & $51.8\pm0.9$ & $\gtrsim14.77$ & $\leq1.0$ & bl,bl-3 \\
 &  &  & \nv\ 1239 & 4782.59 & \nv\ 1243 & 4797.97 & $95.2\pm27.8$ & $14.64\pm0.11$ & $0.66\pm0.08$ & bl-4,6,7\\ 
 & & & \siiv\ 1394 & 5380.90 & \siiv\ 1403 & 5415.53 & $39.1\pm11.9$ & $14.00\pm0.24$ & $0.41\pm0.09$ & bl-9\\ 
 &  &  & \civ\ 1548 & 5976.96 & \civ\ 1551 & 5986.90 & $240.5\pm26.5$ & $14.72\pm0.21$ & $0.50\pm0.08$ & bl-4,6,7\\
6 & 2.8635 & -2884 & \multicolumn{4}{c}{Lyman Series} & $27.8\pm0.6$ & $15.78\pm0.01$ & $1.0$ & \\
 & & & \ovi\ 1032 & 3986.85 & \ovi\ 1038 & 4008.83 & $60.1\pm2.3$ & $14.99\pm0.03$ & $1.0$ & bl-9 \\
 &  &  & \nv\ 1239 & 4786.18 & \nv\ 1243 & 4801.57 & $96.1\pm10.8$ & $15.61\pm0.13$ & $0.59\pm0.04$ & bl-4,5,7\\ 
 & & & \siiv\ 1394 & 5384.95 & \siiv\ 1403 & 5419.60 & $84.6\pm20.8$ & $14.21\pm0.13$ & $0.41\pm0.07$ & bl-3,10\\ 
 &  &  & \civ\ 1548 & 5981.45 & \civ\ 1551 & 5991.40 & $67.7\pm1.8$ & $15.30\pm0.04$ & $0.78\pm0.05$ & bl-5,7\\ 
7 & 2.8671 & -2607 & \multicolumn{4}{c}{Lyman Series} & $111.7\pm3.6$ & $\gtrsim13.81$ & $0.43\sim1.0$ & bl\\
 & & & \ovi\ 1032 & 3990.56 & \ovi\ 1038 & 4012.57 & $126.8\pm2.5$ & $\gtrsim14.90$ & $0.75\sim1.0$ & bl-10 \\
 &  &  & \nv\ 1239 & 4790.64 & \nv\ 1243 & 4806.05 & $120.9\pm1.8$ & $14.87\pm0.04$ & $0.67\pm0.06$ & bl-4,5,6\\ 
 &  &  & \civ\ 1548 & 5987.02 & \civ\ 1551 & 5996.98 & $89.7\pm1.5$ & $14.58\pm0.05$ & $0.67\pm0.07$ & bl-5,6\\ 
8 & 2.8838 & -1323 & \multicolumn{4}{c}{Lyman Series} & $72.3\pm3.8$ & $\gtrsim13.51$ & $0.31\sim1.0$ & bl\\
 & & & \ovi\ 1032 & 4007.79 & \ovi\ 1038 & 4029.90 & $52.6\pm2.1$ & $\gtrsim14.48$ & $0.75\sim1.0$ & bl,bl-5,9 \\
 &  &  & \nv\ 1239 & 4811.33 & \nv\ 1243 & 4826.80 & $55.6\pm1.7$ & $14.30\pm0.10$ & $0.48\pm0.12$ & bl-9,10\\ 
 & & & \siiv\ 1394 & 5413.24 & \siiv\ 1403 & 5448.08 & $38.6\pm1.4$ & $13.42\pm0.17$ & $0.46\pm0.18$ & \\ 
 &  &  & \civ\ 1548 & 6012.88 & \civ\ 1551 & 6022.88 & $56.1\pm5.5$ & $14.19\pm0.11$ & $0.59\pm0.11$ & bl-11\\ 
9 & 2.8859 & -1161 & \multicolumn{4}{c}{Lyman Series} & $87.2\pm1.3$ & $\gtrsim13.73$ & $0.39\sim1.0$ & bl\\
 & & & \ovi\ 1032 & 4009.96 & \ovi\ 1038 & 4032.07 & $76.0\pm2.9$ & $\gtrsim14.77$ & $0.90\sim1.0$ & bl,bl-6,10 \\
 &  &  & \nv\ 1239 & 4813.93 & \nv\ 1243 & 4829.41 & $80.8\pm1.5$ & $15.06\pm0.02$ & $0.55\pm0.05$ & bl-8,10\\ 
 & & & \siiv\ 1394 & 5416.17 & \siiv\ 1403 & 5451.02 & $90.1\pm43.3$ & $13.78\pm0.19$ & $0.40\pm0.13$ & bl-5\\ 
 &  &  & \civ\ 1548 & 6016.13 & \civ\ 1551 & 6026.14 & $83.0\pm8.1$ & $14.56\pm0.17$ & $0.59\pm0.07$ & bl-11,12\\ 
10 & 2.8879 & -1007 & \multicolumn{4}{c}{Lyman Series} & $45.2\pm4.2$ & $14.70\pm0.03$ & $1.0$ & bl\\
 & & & \ovi\ 1032 & 4012.03 & \ovi\ 1038 & 4034.15 & $29.0\pm4.0$ & $\gtrsim14.18$ & $0.84\sim1.0$ & bl,bl-7,9,11 \\
 &  &  & \nv\ 1239 & 4816.41 & \nv\ 1243 & 4831.90 & $46.8\pm2.1$ & $14.03\pm0.12$ & $0.47\pm0.18$ & bl-9\\ 
 & & & \siiv\ 1394 & 5418.96 & \siiv\ 1403 & 5453.83 & $18.0\pm13.2$ & $12.56\pm0.57$ & $0.46\pm0.48$ & bl-6\\ 
 &  &  & \civ\ 1548 & 6019.23 & \civ\ 1551 & 6029.24 & $28.6\pm3.4$ & $13.78\pm0.12$ & $0.56\pm0.18$ & bl-12\\ 
11 & 2.8908 & -784 & \multicolumn{4}{c}{Lyman Series} & $47.4\pm1.3$ & $\gtrsim13.41$ & $0.36\sim1.0$ & bl\\
 & & & \ovi\ 1032 & 4015.02 & \ovi\ 1038 & 4037.16 & $64.1\pm6.9$ & $\gtrsim14.12$ & $\leq1.0$ & bl,bl-7,10,12 \\
 &  &  & \nv\ 1239 & 4820.00 & \nv\ 1243 & 4835.50 & $93.3\pm2.1$ & $15.16\pm0.02$ & $0.13\pm0.05$ & bl-12\\ 
 &  &  & \civ\ 1548 & 6023.72 & \civ\ 1551 & 6033.74 & $109.7\pm4.0$ & $14.10\pm0.15$ & $0.56\pm0.16$ & bl-8,9\\ 
12 & 2.8933 & -592 & \multicolumn{4}{c}{Lyman Series} & $132.5\pm6.2$ & $\gtrsim13.44$ & $0.26\sim1.0$ & bl\\
 & & & \ovi\ 1032 & 4017.60 & \ovi\ 1038 & 4039.75 & -- & -- & $\leq1.0$ & bl,bl-11 \\
 &  &  & \nv\ 1239 & 4823.10 & \nv\ 1243 & 4838.61 & $107.6\pm2.1$ & $15.24\pm0.02$ & $0.14\pm0.04$ & bl-11\\ 
 & & & \siiv\ 1394 & 5426.48 & \siiv\ 1403 & 5461.40 & $129.9\pm3.9$ & $14.21\pm0.06$ & $0.13\pm0.01$ & \\ 
 &  &  & \civ\ 1548 & 6027.59 & \civ\ 1551 & 6037.61 & $117.6\pm9.7$ & $14.10\pm0.15$ & $0.52\pm0.17$ & bl-9,10\\ 
\enddata
\end{deluxetable*}

\subsection{Line-locked Signatures \label{sec:lock_result}}

\begin{figure*}[htbp]
\centering
\includegraphics[width=0.95\textwidth]{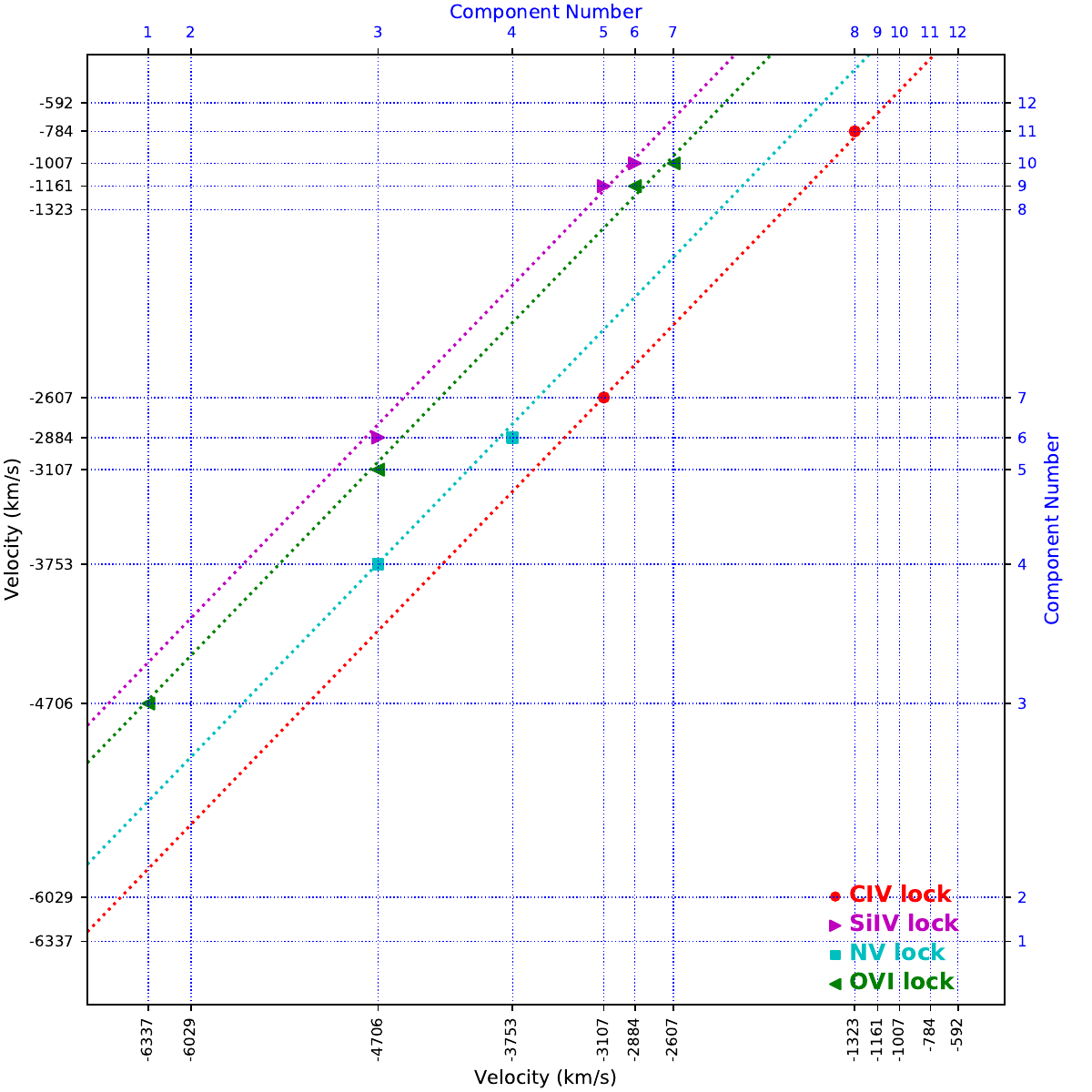}
\caption{The component matrix shows all possible line locks in \civ, \siiv, \nv\ and \ovi\ based on their velocity separations. We consider two systems locked when their velocity separation is a close match (with errors less than 10\%) to their laboratory doublet separation (see text for details). All line-locking pairs corresponding to the same ion doublet sit on the same line, where red line for \civ\ doublet, purple line for \siiv, cyan line for \nv, and green line for \ovi. 
\label{fig:matrix}}
\end{figure*}

\begin{figure*}[htb]
\centering
\includegraphics[angle=270, width=0.9\textwidth]{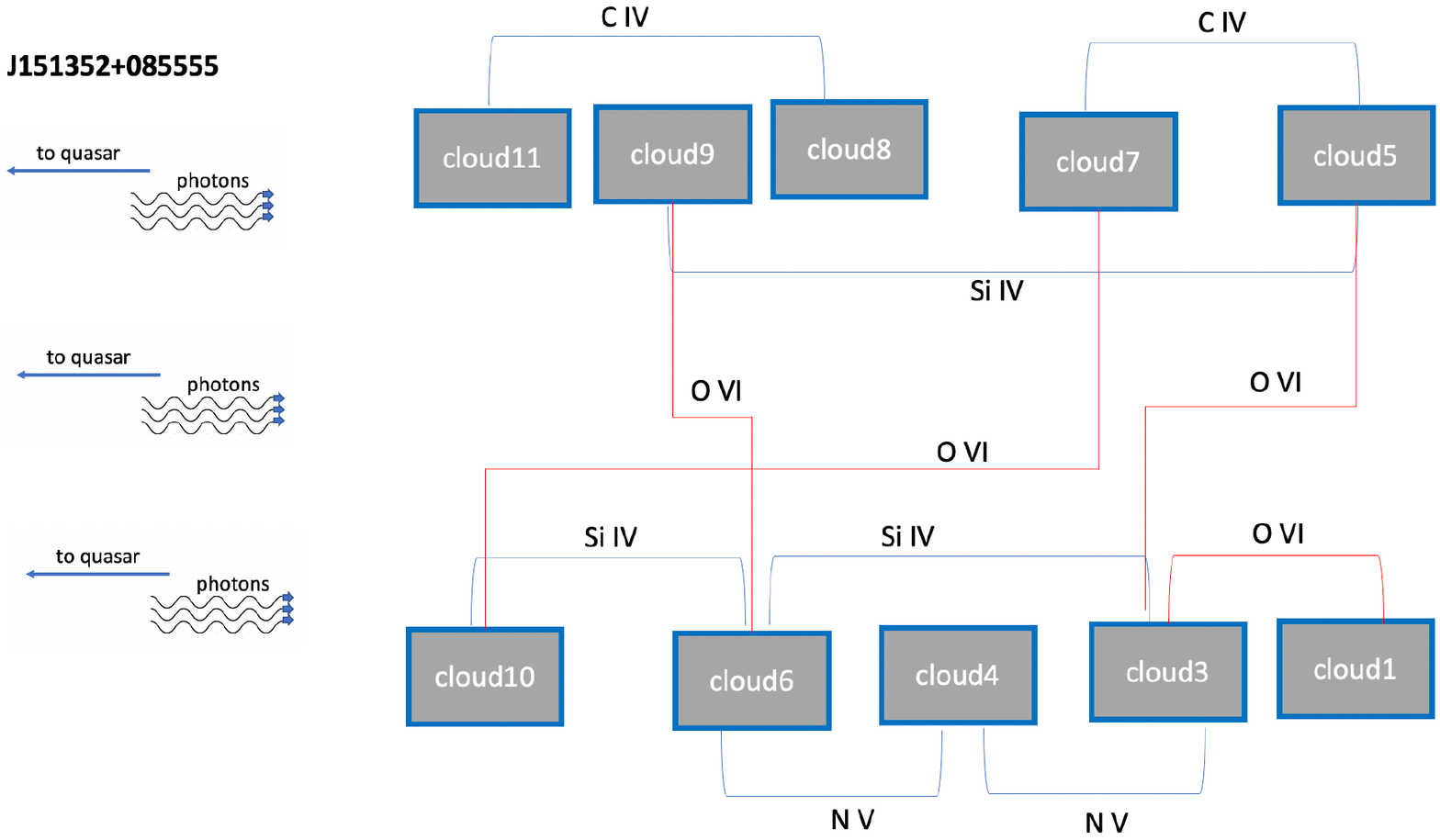}
\caption{The complex absorber systems line-locked through multi-ion doublets detected in the spectrum of \target. All absorbers are exposed to ionizing radiation coming from the quasar on the left. 
The number inside the cloud box is the corresponding component index shown in \Cref{tab:J1513}. We also show if the cloud having \civ, \siiv, \nv\ and \ovi\ doublets identified.}
\label{fig:multi-lock}
\end{figure*}

\begin{figure*}[htbp]
\centering
\includegraphics[width=0.95\textwidth]{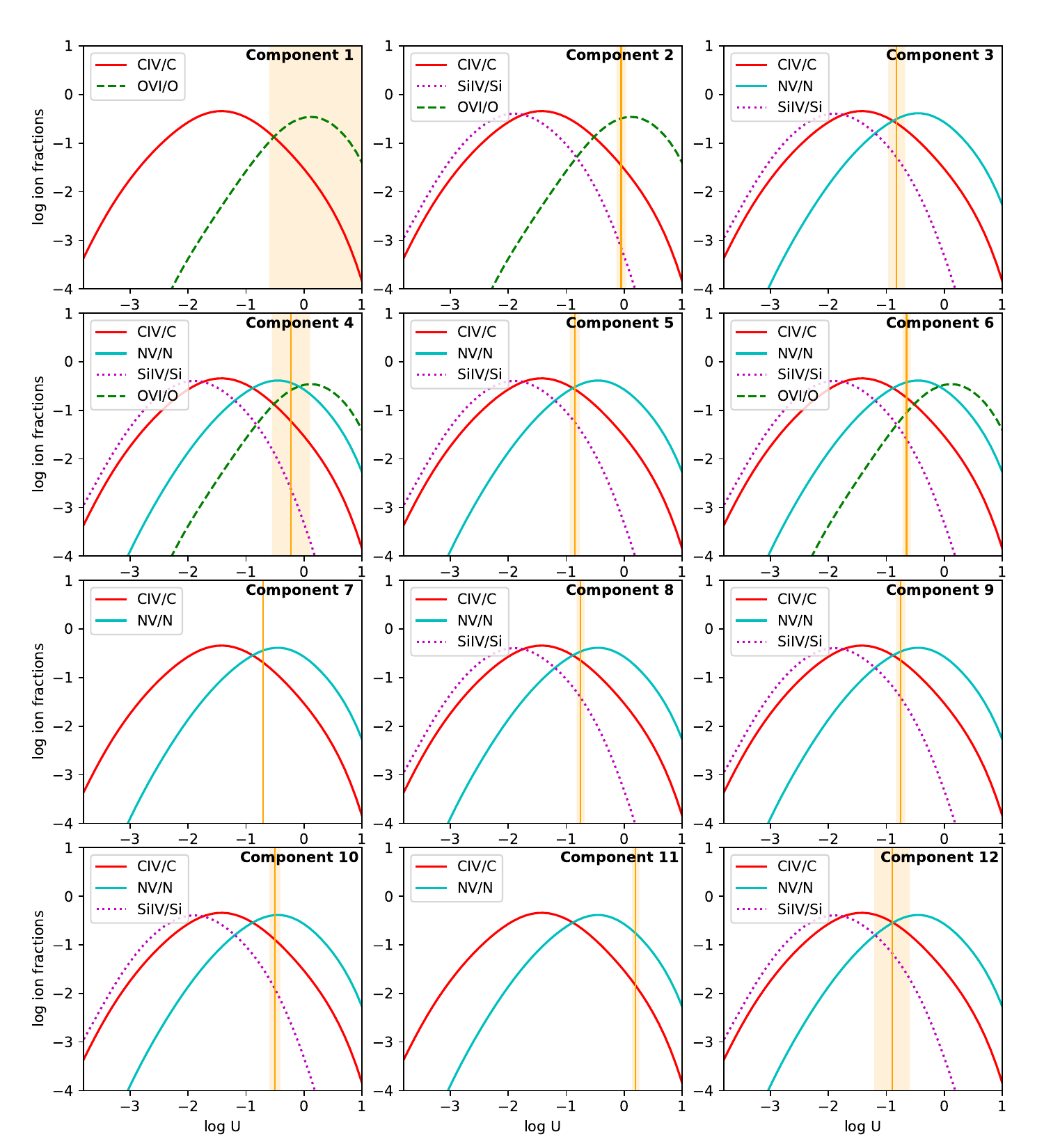}
\caption{Theoretical ionization fractions, $f(M_i)$, for selected stages of the elements Si, C, N and O plotted against ionization parameters $\log U$. The orange vertical lines with orange shadows are the best estimations and errors of $U$ for each component.\label{fig:cloudy}}
\end{figure*}

\begin{figure}
\centering
\includegraphics[width=0.47\textwidth]{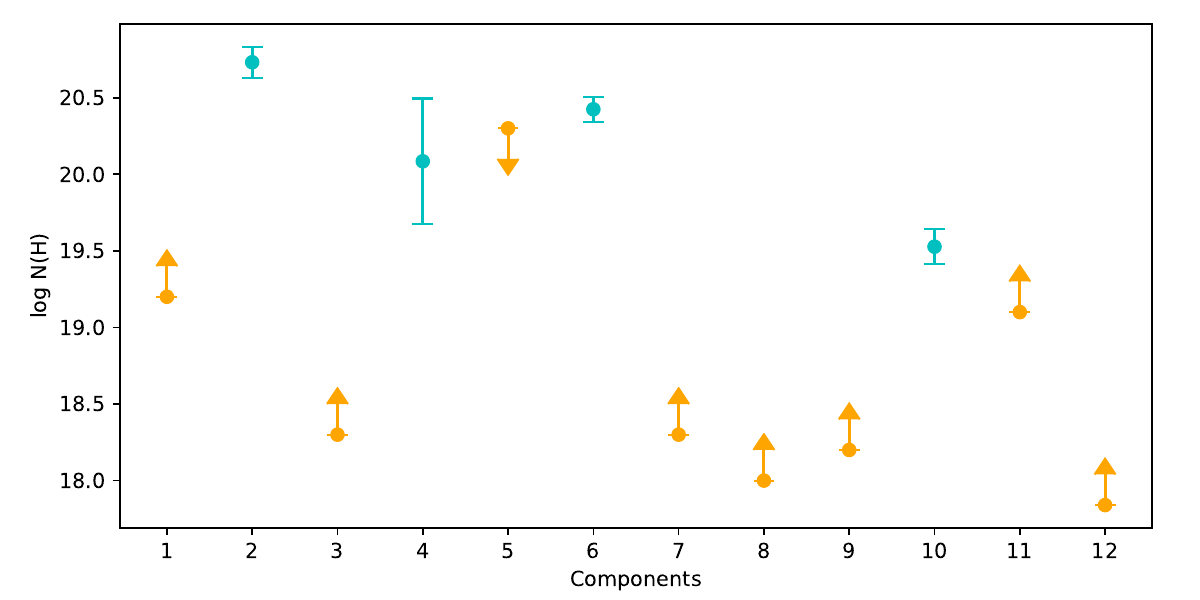}
\caption{Total hydrogen column density log $N$(H) for components 1-12. The orange and cyan color codes correspond to the upper or lower limits and the well-measured column densities, respectively.}
\label{fig:tot_H}
\end{figure}

\begin{figure}
\centering
\includegraphics[width=0.47\textwidth]{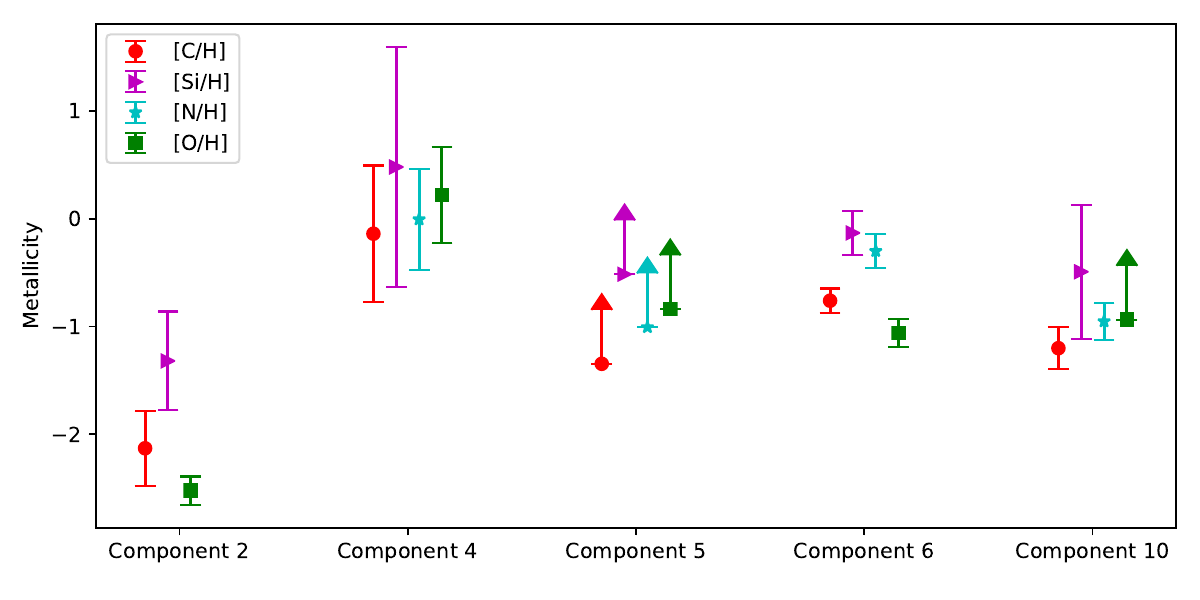}
\caption{C, Si, N, O abundances for components 2, 4, 5, 6 and 10.}
\label{fig:metallicity}
\end{figure}

We searched for line-locked pairs of absorption lines corresponding to the \civ, \siiv, \ovi, and \nv\ doublets in the quasar \target\ spectrum, utilizing the component matrix in velocity space displayed in \Cref{fig:matrix}. Line-locked pairs can be identified within the matrix by drawing a straight line with a slope of $+1$, with the intercept corresponding to the doublet velocity separation, which is approximately $498~\rm{km~s^{-1}}$ for \civ, $964~\rm{km~s^{-1}}$ for \nv, $1939~\rm{km~s^{-1}}$ for \siiv, and $1650~\rm{km~s^{-1}}$ for \ovi. 

We define two absorbers as line-locked if their velocity difference meets the criterion:
\begin{equation}
    \mid{\delta v-v_{\rm sep}} \mid \le \min(\rm{b_1}, \rm{b_2}, 0.1\times v_{\rm sep}),
\end{equation}
where $v_{\rm sep}$ represents the velocity separation of the doublet, and $\rm{b_1}$ and $\rm{b_2}$ are the Doppler parameters of the two lines, respectively. Confirmed line-locked pairs are indicated using colored brackets in \Cref{fig:civ,fig:siiv,fig:nv,fig:ovi}. 

In \Cref{fig:civ}, two line-locked pairs of \civ\ absorbers are identified: pairs (5, 7) and (8, 11), with separations matching the \civ\ $\lambda1548, 1551$ doublet. We also detect line-locking among other ionic doublets, including \siiv\ $\lambda1393, 1403$, \nv\ $\lambda1239, 1243$, and \ovi\ $\lambda1032, 1038$. For \siiv, two line-locked groups are found: (3, 6, 10) and (5, 9), as marked in red in \Cref{fig:siiv}. In the case of \nv, three successive doublets (3, 4, 6) form a line-locked complex, highlighted in \Cref{fig:nv}. We also identify three line-locked pairs in the \ovi\ doublet: (1, 3, 5), (6, 9), and (7, 10). 

The line-locking web can manifests as velocity-stabilized absorption features in quasar spectra, caused by the combined effects of multiple line-locking processes involving ions like \civ, \nv, \siiv, and \ovi, etc. 
Clouds in the web show distinct velocity separations matching the doublet spacings of common UV resonance transitions. 
From the component matrix in \Cref{fig:matrix}, we  have identified one of the largest line-locking web known to date.  
This web contains ten absorbers, which are interconnected through doublets of \civ, \siiv, \nv, and \ovi. 
Its structure is illustrated in the schematic shown in \Cref{fig:multi-lock}. 
Such web configuration could occur by chance, we estimated the probability of alignment by chance following the approach of \citet{Srianand02}. 
We randomly populated 12 absorbers within the velocity space of -7000 to 0~km/s. 
The probability of detecting the arrangement depicted in \Cref{fig:multi-lock} is found to be $< 10^{-8}$. 
This low chance alignment probability, together with the observation of multiple line-locked doublets in the same quasar \citep[e.g.,][]{Ganguly08, Hamann11}, supports line-locking as a natural consequence of radiative acceleration \citep{Srianand02, Juranova24}. 
Traditionally, theoretical studies of line-locking have often employed simplified two-cloud models to explore the phenomenon \citep{Lewis23}. 
The identification of the largest line-locking web calls for more sophisticated modeling.
Potential interactions among the line-locked clouds in the web presents an interesting research topic for the future.

\subsection{Photoionization model \label{sec:photo}}
In this section, we explore the physical conditions of the absorber systems based on photoionization modeling. To gain insights into the ionization states, total column density, and element abundances of the absorbers, we employed the \cloudy\ 2023 release \citep{Chatzikos23}. Although a specific ionic column density can arise under various physical conditions, combining constraints from multiple ionic species helps narrow down the range of possible gas properties in the absorption systems. All 12 narrow absorption line (NAL) components identified (components 1 to 12) in \Cref{sec:lock_result} have measured column densities for at least two ion species, as listed in \Cref{tab:J1513}. 

We run \cloudy\ using a generic fixed \hi\ column density of $\log N(\hi)$(\cmN)~$=15$ that ensures that the clouds are optically thin in the Lyman continuum. This is justified by the measurements and upper limits on $N(\hi )$ that we derive from the data (\Cref{tab:J1513}). In this optically thin regime, the column density ratios needed for our analysis do not depend on $N(\hi)$ nor $N$(H). The calculations assume solar abundances and a standard quasar ionizing spectrum. 

The ionization state of the absorbers is described by the ionization parameter, $U$, which is the dimensionless ratio of the number density of hydrogen-ionizing photons at the illuminated face of the clouds to the number density of hydrogen atoms. \Cref{fig:cloudy} shows the calculated ionization fractions in important ions compared to well-measured column density ratios in the data (e.g., \siiv/\civ, \civ/\nv, \nv/\ovi, etc.). These comparisons yield estimates of the ionization parameter $U$ (shown by the orange vertical lines in the figure) with uncertainties (orange shadows) based on the column density uncertainties listed in the data tables (see Section 4.3 in \citet{chen18} and Section 4.1.1 in \citet{Chen19} for more details). In particular, if a component exhibiting absorption lines for multiple ions (e.g. \siiv, \civ, \nv); in that case, the ion ratios (e.g. \siiv/\civ, \civ/\nv, \nv/\siiv) can yield distinct $U$ values. We derive a weighted mean (based on column densities uncertainties) to be the best $U$ value and a weighted error to be the measurement error.

Subsequently, we derive the total hydrogen column density, $N(\rm{H})$, for components 2, 4, 6 and 10 that have well-measured $N$(\hi ) values, and estimate upper or lower limits on $N$(H) for other components with only upper or lower limits on $N$(\hi ). This is accomplished by applying an ionization correction, \hi /H, appropriate for the derived $U$ values to the $N$(\hi ) values measured from the data. The results are shown in \Cref{fig:tot_H}. 
The derived values span ranges of $\log N$(H) (\cmN) from $\sim18$ to $\sim20.5$ and $\log U$ from $\sim-1$ to $\sim0$, consistent with the findings of \citet{Srianand02}. In that study, the physical conditions of a single component (component e, corresponding to our component 5) were explored, yielding $-2.0 \leq \log U \leq -0.6$ and $10^{19}$ \cmN $\leq N(\textrm{H}) \leq 10^{20}$ \cmN, in agreement with our results of $\log U \sim -0.9$ and $N(\textrm{H}) \lesssim 10^{20.3}$ \cmN. 

We then estimate the component metallicities using the general relation \citep[see more details in][]{chen18, Chen19}:
\begin{equation}
\left[\frac{\mbox{M}}{\mbox{H}}\right]=\log\left(\frac{N(\mbox{M}_i)}{N(\hi)}\right)+\log\left(\frac{f(\hi)}{f(\mbox{M}_i)}\right)+\log\left(\frac{\mbox{H}}{\mbox{M}}\right)_{\odot},
\end{equation}
where $(\mbox{H/M})_{\odot}$ is the solar abundance ratio of hydrogen to some metal $\mbox{M}$, $N(\hi)$ and $f(\hi)$ are the column density and ionization fraction of \hi, respectively, and $N(\mbox{M}_i)$ and $f(\mbox{M}_i)$ are the column density and ionization fraction of ion $\mbox{M}_i$ of metal $\rm{M}$, respectively. 
In particular, $N(\mbox{M}_i)$ and $N(\hi)$ are derived from our fitting process. We use our estimates of $U$ to determine the ionization correction factor $f(\hi )$/$f(\rm{M}_i)$ from the \cloudy\ calculations, and then plug in measured values of the column densities to obtain the metal abundance [M/H].

The results for [Si/H], [C/H], [N/H] and/or [O/H], depending on lines available, are shown in \Cref{fig:metallicity}. The error bars shown in this figure are derive only from the measurement uncertainties in the corresponding ion column densities. 
\Cref{fig:metallicity} shows a mixture of metallicity results for the five components. We have metallicity estimates for components 2, 4, 5, 6, and 10 because their \hi\ column densities have been accurately measured or an upper limit has been estimated. For the other components, we have not displayed the limits because the lower limits on $N$(\hi) result in high upper bounds for metallicity, well above solar, which do not offer meaningful constraints for our analysis.

Component 2 shows substantially sub-solar metallicity, very high velocity, and a sharp line profile, which are very different from the other 10 locked components. 
This indicates component 2 may form in a physically distinct region compared to the other locked components, likely farther away from the background continuum source-perhaps in the halo. This difference could explain why it is not locked with the other components.
In components 6, we also find evidence for enhanced nitrogen abundance, with [N/C]~$>0$, as reported in the literature \citep{Petitjean94,Petitjean08,Fechner09}. 
For example, \citet{Fechner09} find 27$\%$ of the associated systems show clearly enhanced nitrogen. 
In Paper I, we also obtained reasonable [N/C] value for component 9 in Quasar~J221531-174408.
Using data from this study (component 4, 5, 6, 10) and Paper I (component 9), we find 20$\%$ (1 out of 5) NALs show evidence of enhanced nitrogen. 
Due to the small sample size of 5, this result should be considered preliminary and warrants further investigation with larger datasets to confirm its validity.

Assuming a tenth solar metallicity, cloud sizes ranging from 0.1 to 1~pc, and hydrogen column densities of $\log N(\textrm{H})$(\cmN) ranging from 18 to 20.5 as derived from \cloudy, we estimate the masses of typical absorbers to lie between $1 \times 10^{-4}$ and $3$~$M_\odot$.
Following the method of Paper I, we also estimate the total kinetic energy of the outflow system identified through line-locking, and evaluate its impact on the host galaxy's evolution from the energetic perspective. 
Assuming typical distances of $\sim$10–100~pc for the line-locking clouds, we obtain kinetic energies $K \lesssim 1.65-165 \times 10^{53}$~ergs, and total kinetic luminosity of $\dot{K} \lesssim (0.2-2) \times 10^{43}$~ergs~s$^{-1}$ for this line-locking web. 
The resulting ratios between the kinetic energy rate and bolometric luminosity, $\dot{K}/L_{\rm Bol} \lesssim (0.5-5) \times 10^{-5}$, is two orders of magnitude lower than the threshold of 0.005 to 0.05 required for effective feedback to the host galaxies \citep{Scannapieco04, DiMatteo05, Prochaska09, Hopkins10}.
The conclusion is that the kinetic luminosity of the NAL outflow system is quite low compared to the bolometric luminosity of the quasar, which is consistent with the findings for the NAL outflow system in Quasar~J221531-174408, as reported in our Paper I.

\begin{figure}[htbp]
\centering
\includegraphics[width=0.48\textwidth]{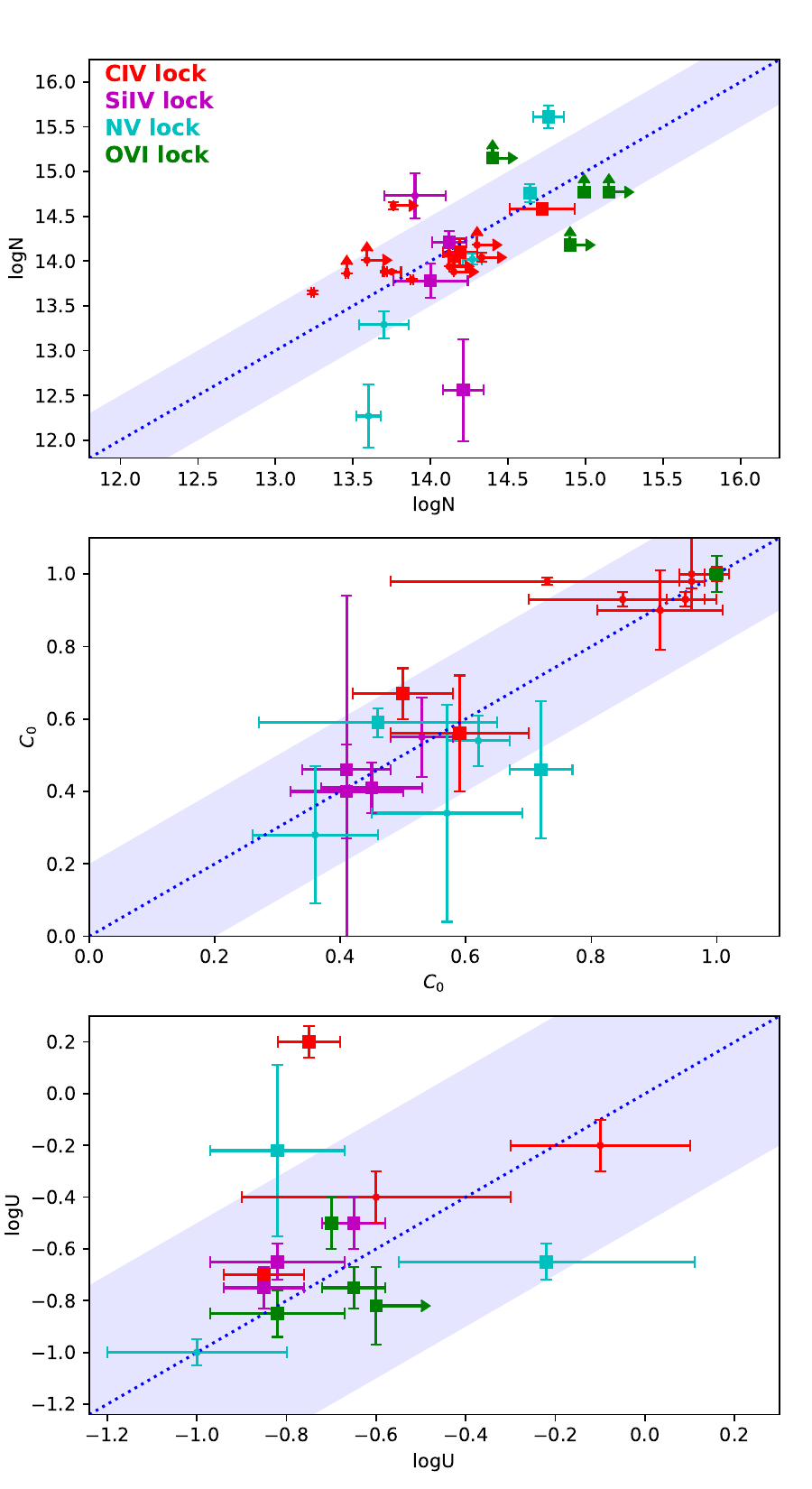}
\caption{Comparison between the derived properties of the shadowed cloud (X-axis) and shadowing cloud (Y-axis) in the same line-lock pair. 
Upper panel: logarithm of column density. Middle panel: covering fraction. Lower panel: logarithm of ionization.
The blue dashed line indicates a 1:1 ratio. In the upper and lower panels, the blue shadows delineate the region where the parameters of the shadowing clouds are within a factor of $\sim3$ above and below those of the shadowed clouds. In the middle panel, the blue shadow marks the 20\%\ uncertainty area around the 1:1 ratio line.
The sample contains all the line-locked systems that were analyzed in \citet{chen18,Chen19,Chen24} (represented by filled dot symbols) and this paper (represented by the filled square symbols).}
\label{fig:N_U_C}
\end{figure}

\section{Discussion}
\subsection{Comparison with Previous Studies}
Quasar \target\ is one of only four known quasars exhibiting line-locking (LL) signatures in at least \civ, \siiv, and \nv\ doublets simultaneously. The other three are Q1303+308 \citep{Turnshek84a, Foltz87, Braun89}, J221531-174408 \citep{Chen24}, and SDSS~J092345+512710 \citep{Lin20b}. 
As illustrated in Figure~\ref{fig:multi-lock}, quasar \target\ exhibits one of the most complex LL structures reported in the literature. Ten out of the twelve identified NAL absorbers are linked through radiative forces, as evidenced by their velocity separations corresponding to the \civ, \siiv, \ovi, and \nv\ doublets.
As discussed in Section~2, \citet{Srianand02} is the first to identify the LL phenomena in quasar~\target. They have investigated its LL properties using the UVES observation obtained at year 2000. Our analysis here uses the combined data obtained from UVES observation at year 2000, 2002, and 2003, and reveals some differences compared to the findings of \citet{Srianand02}. 

While \citet{Srianand02} focused on LL signatures associated with the \ovi, \nv, and \civ\ doublet splittings, we also identify two additional LL pairs corresponding to \siiv. 
Both studies have detected two pairs of LLs in \civ; however, we identify one different pair compared to their results. The first \civ\ LL pair reported by \citet{Srianand02} (the b-c pair) is considered marginal in our analysis due to poor matching between the line profiles of the two components. The detected \nv\ LL pairs are consistent between both works.
Our study find two double LL pairs and one triple LL pair in \ovi, while \citet{Srianand02} detected only one triple LL pair. The discrepancies arise mainly from differences in data reduction methods and LL identification technique. 
Specifically, we have identified seven additional LL pairs compared to \citet{Srianand02}, thanks to the improved data reduction provied by \citet{Murphy19} and the use of a different LL signature identification algorithm. 

\citet{Srianand02} have studied physical parameters of component e (corresponding to component 5 in this study) using a photoionization model. They reported a covering fraction of 0.4 and column density $\log N(H \mathrm{I})$ around 16.3, while we obtained values of $\gtrsim0.4$ and $\lesssim15.8$, respectively. They have shown the ionization parameter of component 5 should be in the range $-2.0 < \log U < -0.6$, while we gave a value around -0.9. For the total hydrogen column density of component 5, \citet{Srianand02} gave $19 \lesssim \log N(H) \lesssim 20$, and we obtained a value of $\log N(H) \lesssim 20.3$. For all the other components presented in their Table~1, the covering fractions generally agree with results presented in our \Cref{tab:J1513}. 
In \citet{Srianand02} they argued the distance of the absorbers to the background continuum source is between 0.1~pc to a few kpc, while we simply adopted typical values of 10-100~pc when estimating the total gas mass in the outflow in \Cref{sec:photo}. 
We also placed constraints on the metallicities of five components to be equal to or less than solar, whereas \citet{Srianand02} did not report any metallicity measurements.
The above comparisons focusing on component 5 suggest both studies yielded similar line fitting and ionization results. 
And we can conclude that the main differences between the two studies are the LL structure identified, possibly due to different LL identification technique.

\subsection{`Shadowed' vs `Shadowing' absorbers}

We find ten out of the twelve NALs identified in the \target\ spectra appear line-locked, which represents a notably high percentage.
This suggests that line-locking may be prevalent in quasar outflow, like suggested by \citet{Bowler14}, and \citet{Chen21}.
Although the specific physical mechanism is still under debate \citep[see the discussion in][]{Bowler14}, it is widely agreed that line-locked systems are related to the physical process of radiative acceleration \citep{Milne26, Scargle73, Braun89, Lewis23, Dannen24}, where successive shielding of the outflowing clouds locks the clouds one by one in outflow velocity.
Using simulations, \citet{Lewis23} suggests the clouds' properties need to be fine-tuned to within 1\% for the clouds to be line-locked. 
The case presented here, together with our previously studied cases \citep{chen18,Chen19,Chen24}, can be used for a preliminary statistical test of this scenario.
We have compared the ion column density, coverage fraction, and $\log U$ between the `shadowed' absorbers and `shadowing' absorbers in \Cref{fig:N_U_C}. 
We find that most ion column densities and ionization parameters of the shadowing clouds lie within a factor of $\sim3$ of those of the shadowed clouds.
Additionally, the coverage fractions of the shadowing clouds are comparable to those of the shadowed clouds.
This generally supports the study of \citet{Lewis23}, where they argue the two absorbers need to have similar physical properties. 
However, we also find exceptions, suggesting that clouds with quite different physical conditions can also be locked together, and fine-tuning is not always required. 

One possible solution to the discrepancies between observational evidence and the theoretical prediction of \citet{Lewis23} is to add more clouds ($>10$) and multi-ionic LL ($>3$) to the modeling process, instead of using a simple two clouds one doublet model.
Further theoretical modeling and detailed simulations are needed to reveal the mechanisms behind these line-locking phenomena.

\section{Conclusion}
As the second paper in a series of TOLL project, we have conducted a detailed investigation of line-locking signatures in QSO~\target\ using high-resolution, high signal-to-noise ratio archived VLT-UVES spectra. 
Our study provides a significant improvement in sensitivity to weak and narrow associated absorption lines compared to previous outflow line-locking surveys conducted with medium-resolution spectra, such as those from the SDSS. 
A total of 12 associated absorption line systems were identified, and we performed detailed fitting for each \civ, \nv, \siiv, and \ovi\ narrow absorption line to derive physical properties, including velocity shifts, column densities, and line-of-sight covering fractions, $C_0$. 
We find that these AALs have covering fractions ranging from $\sim$0.4 to $\sim$1, and $\log U$ values ranging from -1.0 to 0.0. 
Ten out of the 12 identified AALs exhibit line-locking signatures, forming one of the largest line-locking web known to date.
Discovery of this line-locking web calls for more sophisticated theoretical model and simulation works to explain its formation and evolution.

We have also compared the properties of the shadowed and shadowing cloud in the same line-locking pair, using data from our previous works and this study. 
We find they tend to have similar $\log N$, $\log U$, and $C_0$ values, with a few exceptions. 
Our findings suggest that it is necessary to use high-resolution, high-SNR spectra to study the line-locking phenomena in quasar outflow.
The data presented in this paper can be used to test theoretical models in the future.

\section*{acknowledgments}
 
We want to extend our heartfelt gratitude to Prof. Fred Hamann, now retired, for initiating this project and for his invaluable guidance as the PhD advisor of Dr. Chen Chen. 
To enjoy a peaceful retirement life, Prof. Fred Hamann has opted to remove his name from the co-author list. His mentorship and expertise have been instrumental in shaping the direction of this project. 
We wish him a happy and healthy retirement. 
We also thank Prof. Michael Murphy for providing the quasar spectrum. 
This work was supported in part by the National Natural Science Foundation of China (12103097, 12073092), the National Key R\&D Program of China (2020YFC2201400). CC acknowledges the support of the `Three Levels' Talent Construction Project of Zhuhai College of Science and Technology. Z. C. He acknowledges the support of the National Natural Science Foundation of China (Grant Nos. 12222304, 12192220, and 12192221). 

\bibliography{reference}
\bibliographystyle{aasjournal}

\end{document}